# Ultra-compact plexcitonic electro-absorption modulator

Ruoyu Yuan, Jason Lynch and Deep Jariwala[*]

*Department of Electrical and Systems Engineering, University of Pennsylvania, PA, 19104, Philadelphia, USA*

*\* Corresponding author: dmj@seas.upenn.edu*

**Abstract:** Compact electro-optic (EO) modulators with large extinction ratios, low-switching energies, and high operation speeds are desirable for integrated photonic and linear optical computing. Traditional 3D semiconductors and dielectrics are unsuitable for achieving such modulators due to the small magnitude of EO effects in them. Excitonic 2D semiconductors present a unique opportunity in this regard given their large and tunable optical constants near the excitonic resonances. However, strategies for confining and electrically tuning the excitons into compact EO modulators haven't been realized thus far. Here, we design and simulate an ultra-compact, plexcitonic (strongly-coupled exciton-plasmon) electro-absorption modulator (EAM) with a sub-micron linear footprint operating close to the excitonic peak of the $WS_2$ monolayer (641 nm) hybridized with the plasmon mode of a silver slot waveguide. Electrostatically injected free carriers in $WS_2$ modulate the light-matter interaction via Coulomb screening of the excitons as well as promoting the formation of charged excitons (trions). For our optimized designs, the EAM is expected to achieve a 9.1 dB extinction ratio, concurrently with a 7.6 dB insertion loss in a 400 nm lateral footprint operating with a predicted < 3 fJ/bit switching energy at 15 GHz for 3-dB bandwidth modulation. Our work shows the potential of plexcitonic quasi-particles for integrated optical modulators.

**Introduction**

In the field of optical communications, electro-absorption modulators (EAMs) are a type of device used to transmit digital data in the non-return-to-zero on-off keying format. An ideal EAM for on-chip integration should have a large extinction ratio, low insertion loss, low switching energy, and a small footprint. While it is possible to adjust the extinction ratio (ER) and insertion loss (IL) by adjusting the length of the EAM, the ratio of attenuation coefficients between the "Off-state" and the "On-state" should be greater than 2 for high tunability and low-loss [1]. In order to be energy-efficient compared to electric interconnects, the switching energy of an EAM should be less than 1pJ/bit [2]. However, traditional silicon-based EAMs have relatively large footprints due to weak electro-optical effects [3], resulting in high energy consumption and limiting their potential for integration. Researchers have focused on hybrid silicon or SiNx optical modulators with different active layers, such as quantum wells, quantum dots, graphene, indium tin oxide (ITO), and transition metal dichalcogenides (TMDs) [4] , in an effort to shrink the devices to submicron lengths [5–12]. However, this approach still has resulted in micron modulation lengths since silicon waveguides do not support modes with extremely small cross-sections.

Therefore, in order to miniaturize EAMs, two things are critical: the adoption of novel, low-dimensional materials that support the submicron confinement of light [13–17] and the development of strategies for tuning these light-matter interactions [18]. TMDs such as tungsten disulfide, have highly tunable excitons and support the formation of polaritons that confine light on the nanometers scale, making them suitable for use in EAMs. Plasmonic waveguides, which utilize metals, can confine light at the interface between a metal and a dielectric [19], allowing for strong coupling between photons and free carriers in the metal and breaking the diffraction limit for light confinement. When plasmonic waveguides are coupled to an exciton, a plexciton is formed with the characteristically small mode volume of the plasmon and the tunability of the exciton. In this paper, we propose and simulate an ultra-compact, plexcitonic EAM utilizing the excitonic properties of a $WS_2$ monolayer structured into a multi-layer superlattice structure with alternating layers of $Al_2O_3$ dielectric [20] and the plasmon mode of an Ag slot waveguide to form and control a plexciton. By injecting free carriers into the $WS_2$ and modulating the coupling through dielectric screening of the excitons, charged excitons (trions) formation and enhanced Coulomb scattering [21,22], our optimized design can achieve a 9.1 dB extinction ratio, a 7.6 dB insertion loss, and a < 2 fJ/bit switching energy at 15 GHz for 3-dB bandwidth modulation, all in a 400 nm linear footprint. Our work demonstrates the potential of plexcitons for integrated optical modulators.

**Device structure**

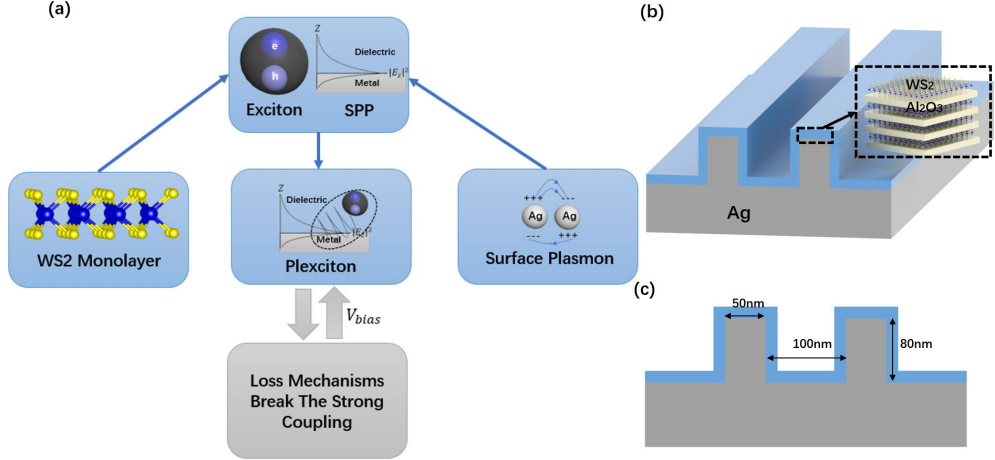

Fig. 1. (a) Mechanisms of the EAM. The strong coupling of excitons and surface plasmon polaritons (SPP) form quasi-particles named plexcitons. Injecting carriers through the biased voltage will turn the strong coupling into the weak coupling. (b) 3D view of the EAM. The slot-shaped plasmonic structure is made of silver with a four-unit superlattice patterned on top (the blue part represents the superlattice). (c) Cross-section view of the EAM with optimized sizes.

Figure 1(a) shows the mechanism of our submicron-scale EAM. In our design, a silver substrate is used to support surface plasmon polariton propagation mode in the visible range that are then coupled to excitons in the $WS_2/Al_2O_3$ superlattice placed top [23]. When the plasmons are engineered to have a similar energy to the exciton, the two states strongly couple to one another to create two distinct energy states where one state has an energy higher than the undisturbed exciton and plasmon and the other state has a lower energy. These two hybridized states are part-exciton and part-plasmon called plexcitons. Strong coupling occurs when energy is exchanged between the excitonic and plasmonic states faster than either quasiparticle decays which is expressed as [24,25]

$$g^2 \geq \frac{(\gamma_C - \gamma_X)^2}{16} \quad (1)$$

where $g$ is the coupling parameter which describes the rate at which energy is exchanged between the exciton and plasmon and $\gamma_C$ ($\gamma_X$) is the linewidth of the plasmon (exciton). When the loss mechanisms become significant, equation (1) implies no energy splitting and the device goes into the weak coupling regime. The coupling parameter is proportional to the square root of the number of excitons and the oscillator strength over the mode volume, which is given by [24,25]

$$g \propto \sqrt{\frac{Nf}{V_m}} \quad (2)$$

where $N$ is the number of excitons, $f$ is the oscillator strength of the excitons, and $V_m$ is the mode volume of the waveguide. When no bias voltage is applied, strongly coupled-plexcitons form and photons with energy near the exciton are forbidden from propagating, defining the "Off-state" of our EAM. When carriers are injected into the $WS_2/Al_2O_3$ superlattice, the carriers Coulomb screen the electrons and holes reducing the oscillator strength of the exciton. The injected carriers can also interact with the excitons to form trions that are red shifted from the exciton, and the trions are also coupled to the plasmon. Therefore, by injecting carriers into the $WS_2$, the system goes from a two-coupled oscillator system between the exciton and plasmon to a three-coupled oscillator system where the plasmon is coupled to both the exciton and trion, but the exciton and trion do not interact [26,27]. The plasmon can still hybridize with both the exciton and trion so long as it exchanges energy with these states faster than any of the states decay. However, the exciton and trion both have smaller oscillator strengths in number density than the exciton in the "Off-state" which reduces the coupling parameter in accordance with equation (2). Because of the reduced coupling parameter, the system enters the weak coupling regime, and the hybrid plexciton states no longer form. In this case, photons with energies close to the exciton energy can propagate through the waveguide, with some absorptive loss by the exciton and trion, defining the "On-state" of the EAM. The tuning of plexcitons creates an ultra-compact EAM operating near the excitonic wavelength of the WS$_2$ monolayer with high extinction ratio and low switching energy while maintaining a durable insertion loss. As the simplest single-interface plasmonic structure only supports TM mode, the slot-shaped structure shown in figure 1(c) increases the in-plane electric field component and thus increases the efficiency of the coupling between excitons and surface plasmon polaritons.

**Optical dispersion and mode characteristics**

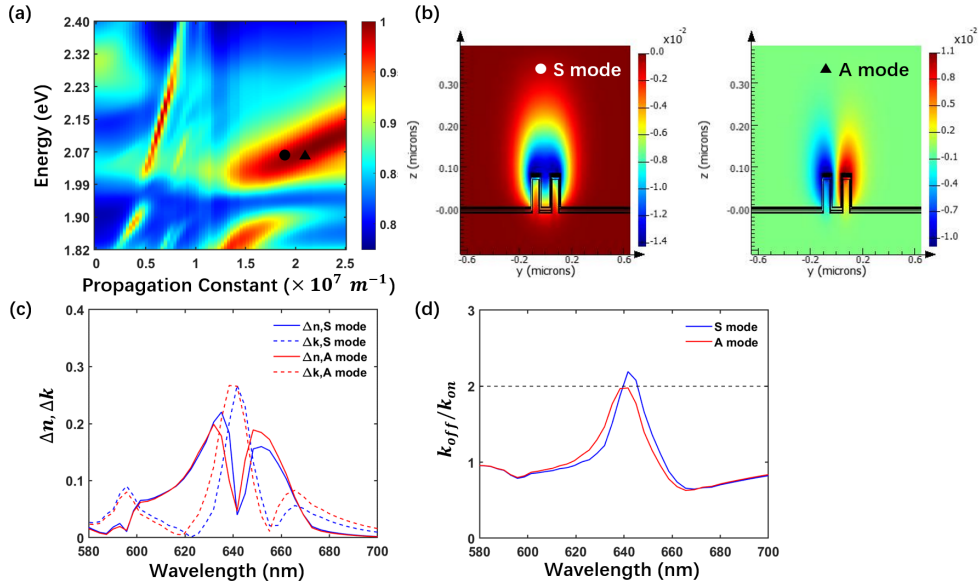

Fig. 2. (a) Dispersion relations of the plasmonic structure through FDTD simulation after logarithmic and normalized processing. Red curves show the SPP modes and the plasmonic mode that are supported by the structure. See Ref. [28] for the method of dispersion calculation. (b) The distribution of the electric field component along the propagation direction for the symmetric mode (S mode) and antisymmetric mode (A mode). (c) Refractive index changes for the S mode and A mode. (d) The figure of merit to describe the performance of the EMA for both modes.

Lumerical MODE and Lumerical Finite-Difference Time-Domain (FDTD) simulations are used to analyze and optimize the optical characteristics of the EAM with a length of 400 nm, using the gate-tunable refractive index data from Ref. [21]. Figure 2 compares two dominant SPP modes and their differences. Figure 2(a) is generated by calculating spectrums for each propagation constant, representing the dispersion relations where a plasmonic mode (parabolic curve) can be seen with small propagation constants (<0.7) and a clear anti-crossing phenomenon around the excitonic peak at larger propagation constants (1.5 to 2) due to strong light-matter coupling. The broadening curve in the upper branch compared to the lower branch is due to background absorption of the WS$_2$ monolayer above its bandgap, and the broadening curve of the SPP mode compared to the plasmonic mode is due to the overlap between two neighboring modes, as shown in Figure 2(b). Figure 2(b) shows the magnitude of the ***Ex*** electric field distribution (***Ex*** points in the direction of SPP propagation). These modes are called the symmetric mode (S mode) and antisymmetric mode (A mode) due to their electric field distributions. Both modes show similar modulation when the refractive index is tuned as shown in Figure 2(c). The maximum change in absorption occurs at 641 nm for both modes, which is defined as the operation wavelength of the modulator. In Figure 2(d), we calculated the figure of merit proposed by Ref. [1], stating that a high-performance EAM should satisfy

$$\frac{k_{off}}{k_{on}} > 2 \tag{3}$$

This criterion guarantees significant modulation while maintaining low insertion loss. In our device, the figure of merit at the operation wavelength (641 nm) for the S mode meets the requirement of a qualified EAM. The S mode and A mode are independent with each other as they are the direct solutions of the eigenmode expansion method (EME). Therefore, the S mode is more suitable for operating our EAM than the A mode, and it should be individually excited in practice to maximize performance.

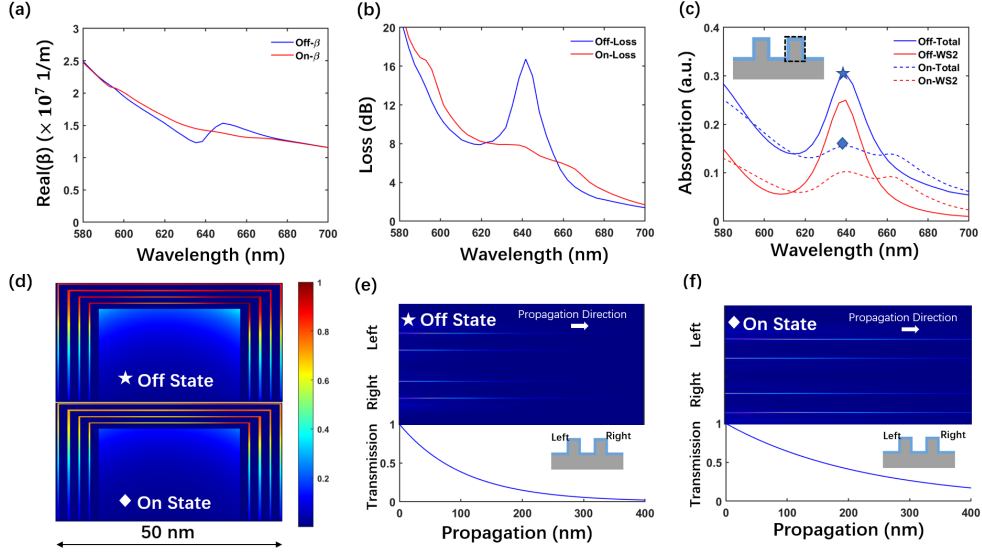

Fig. 3. (a) The dispersion relations of the S mode for the on-state and off-state. (b) The S-mode loss for on-state transmission and off-state transmission of the EAM (400nm linear footprint). (c) The total absorption and the absorption of WS$_2$ layers for on-state transmission and off-state transmission of the EAM. The absorption is calculated within the region of the black-dotted rectangular. (d) Cross-section view of the absorption distribution for the on-state and off-state corresponding to the rectangular part of figure 3(c). The absorption is normalized. (e) Top-down view of the electric field intensity for off-state transmission. (f) Top-down view of the electric field intensity for on-state transmission.

In Figure 3, we continue to study the tunability of the S mode. When no voltage is applied to the structure, the plexciton dispersion relation calculated by the eigenmode expansion method (Figure 3(a)) is consistent with the results obtained using the finite-difference time-domain method (Figure 2(a)). Figure 3(a) shows that the propagation constant reaches its minimum value exactly at the excitonic peak (636 nm), indicating that light cannot propagate through the EAM at this wavelength due the strong light-matter interactions. The system is in a weak coupling region with an applied voltage and will not cause an energy splitting since light can propagate more easily. As a result, the dispersion relation should resemble that of a pure surface plasmon polariton, shown as the red curve in Figure 3(a). In Figure 3(b), we also observe that the absorption peak (641 nm) of plexcitons is red-shifted compared to the excitonic peak of excitons in the WS$_2$ monolayer, which is the operating wavelength of the EAM for the best performance. Our EAM operating at 641 nm has an insertion loss of 7.6 dB and an extinction ratio of 9.1 dB. When analyzing the electro-absorption properties of the EAM, we used the formula [29]

$$P_{abs} = -0.5\omega|E|^2 imag(\varepsilon) \tag{4}$$

where $E$ is the electric field and $\varepsilon$ is the permittivity of the material. The results show that the majority of the optical absorption power comes from the WS$_2$ monolayers, with the remaining power absorbed by silver. This demonstrates the importance of strong coupling mechanisms in boosting the tunability of transition metal dichalcogenides (TMDs). Additionally, light is strongly confined at the four top corners of the slot-shaped waveguide at 641 nm and the intensity exponentially decreases along the propagation direction. These findings suggest that monolayers of other TMDs or ultrathin excitonic semiconductors could also lead to significant tunability due to the strong coupling between excitons and polaritons.

**Electrical switching characteristics**
In order to evaluate the energy efficiency and bandwidth of our EAMs, two theoretical models were developed. Figure 4(a) presents an interdigitated capacitor model for injecting carriers into WS$_2$ monolayers. The superlattice structure can be modeled as interdigitated capacitors with Al$_2$O$_3$ as the dielectric. Because each plate of the capacitor is made of the semiconductor rather than metal, the finite density of state in a WS$_2$ monolayer leads to an additional equivalent capacitor in series with a geometric capacitor, known as a quantum capacitor. The quantum capacitance of the WS$_2$ monolayer plate can be calculated using the following formula [30–32]

$$C_q = Se^2 \frac{g_v m^*}{\pi \hbar^2} \tag{5}$$

where $S$ is the average surface area of the WS$_2$ monolayer, $e$ is the elementary charge, $g_v$ is the valley degeneracy factor (equal to 2 for the WS$_2$ monolayer), and $m^*$ is the effective mass (chosen to be 0.4 $m_0$, the electron mass in free space). Figure 4(b) shows the circuit-level equivalent schematic. The lowest capacitor has one side made of silver and the other side made of a WS$_2$ monolayer, while the other three capacitors are made of a WS$_2$ monolayer on each side, resulting in two quantum capacitors in series. The total capacitance is

$$C_{tot} = \frac{3C_g C_q}{C_q + 2C_g} + \frac{C_g C_q}{C_g + C_q} \tag{6}$$

where $C_g$ is the geometric capacitance. The quantum capacitance is estimated to be 0.196 pF and the geometric capacitance is estimated to be 3.38 fF in our EAM with optimized dimensions. In order to calculate the switching energy of our electro-absorption modulator (EAM), we used the following formula

$$E = \frac{1}{2} C_{tot} V^2 \tag{7}$$

where $C_{tot}$ is the total capacitance, and $V$ is the voltage applied between the capacitors. According to a previous study by Ref. [8], a 67 V voltage was applied to a similar structure with a WS$_2$ monolayer on a silicon-on-insulator (SOI) wafer with a 300 nm SiO$_2$ insulator. In order to achieve the same electric field in our superlattice structure with a 3 nm Al$_2$O$_3$ insulator, the equivalent voltage between capacitors can be 0.67 V. This resulted in a switching energy of 2.96 fJ/bit, which is small due to the ultra-compact size of our EAM. In comparison, other low-dimensional material-based EAMs such as graphene-based EAMs have their total switching energy limited by the quantum capacitance, which puts a physical constraint on energy efficiency improvements.

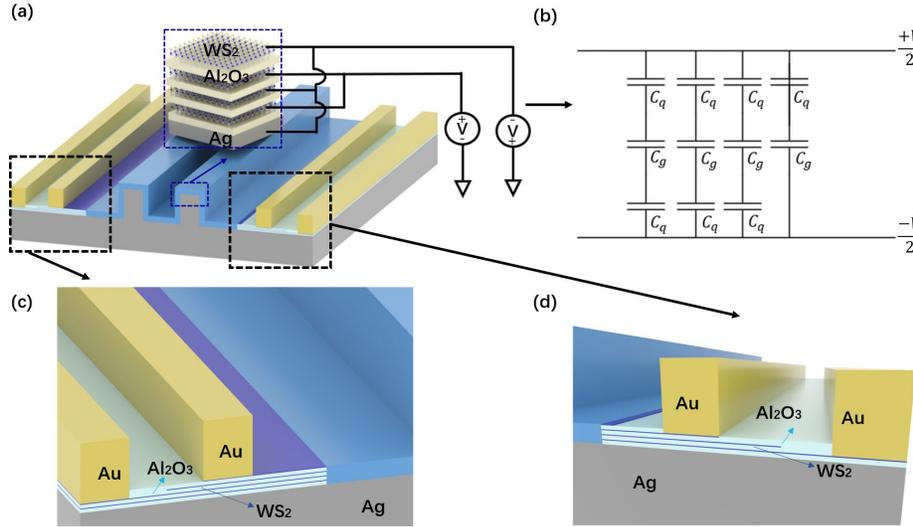

Fig. 4. (a) The Method of applying the voltage to the 4-unit superlattice structures. From the lowest layer to the top layer, the first and the third WS$_2$ monolayer are injected holes, while the second and the fourth are injected electrons. The 'V' in this panel represents the voltage source. (b) The equivalent parasitic capacitance circuit schematic. Both geometric capacitance and quantum capacitance are taken into consideration for a capacitor with semiconductor plates. The voltage across the capacitors has the value $V$, which has the same meaning with the equation (7). (c) The contacts for the second and the fourth WS$_2$ monolayer. The shallow blue part represents the whole superlattice. (d) The contacts for the first and the third WS$_2$ monolayer. Each WS$_2$ monolayer has different width for separately depositing the gold electrode.

Contacts can be created on both sides of the EAM for electrical connections. The right side is designed for contacting the first and third WS$_2$ monolayers, while the left side is for contacting the second and fourth monolayers with a gold electrode. In Figure 4(c), the process involves depositing a 3 nm Al$_2$O$_3$ film, followed by a 0.7 nm WS$_2$ layer. The gold electrode is then patterned on the monolayer. When depositing the second WS$_2$ monolayer, the width should be smaller to avoid contact with the electrode, and the Al$_2$O$_3$ thin film should cover the entire second WS$_2$ monolayer. Similar processes can be used for the third and fourth monolayers and the contacts on the other side of the EAM (Figure 4(d)). A 4-unit superlattice is sufficient for modulation, as more units would require a more complex fabrication process.

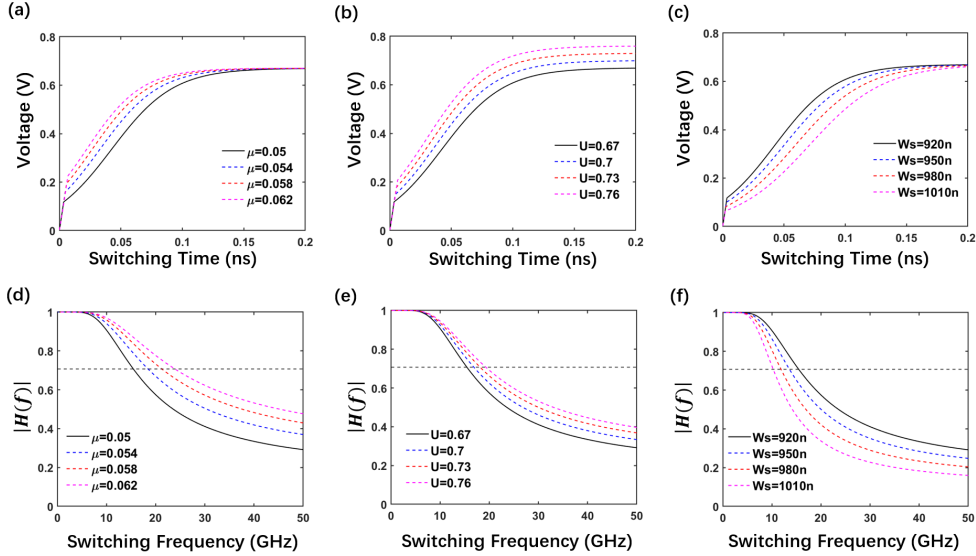

Fig. 5. (a) The voltage between a geometric capacitor vs the switching time of an electric bit under $WS_2$ monolayers with different mobilities. The results with the black correspond to our device. (b) The voltage between a geometric capacitor vs the switching time of an electric bit under $WS_2$ monolayers with different applied voltages. (c) The voltage between a geometric capacitor vs the switching time of an electric bit under $WS_2$ monolayers with different contour widths. (d) The magnitude of the transfer function under different mobilities. (e) The magnitude of the transfer function under different voltages. (f) The magnitude of the transfer function under different contour widths.

The calculation of bandwidth is complex and the simple RC delay model [33] is not effective in estimating materials with relatively high resistivity and significant resistance changes when injecting free carriers, such as $WS_2$ monolayers. Therefore, the rising time to set up the external voltage depends on the minimum time required for free carriers in a $WS_2$ monolayer to be uniformly distributed, as well as the capacitance effect. We have developed a velocity saturation model to simulate the transient performance of our electro-absorption modulator (EAM) (see Supplement 1 for details of the analyses, and the relevant data used in the model is from Ref. [34]). Figure 5 illustrates the quality of $WS_2$ monolayers and the influence of external voltage sources on transient performance. Figures 5 (a), (b), and (c) show different factors that affect the response time of $WS_2$ monolayers to external voltages. For high quality $WS_2$ monolayers, larger mobility leads to faster responses. As carriers are injected from one side to the other, larger mobility means that carriers can quickly occupy all areas of a $WS_2$ monolayer and do not degrade tunability under high frequency. Similarly, decreasing the contour width or decreasing the contact area limits the path that carriers must move through and shortens the rising time. When the external voltage increases, the large electric field accelerates carriers to the saturation velocity within a shorter interval. These results show ways of improving transient performance through careful monolayer growth, higher energy consumption, and a smaller contact area. Among these, the most efficient way is to increase the external voltage, as it is easy to manipulate and paying extra energy also improves optical characteristics such as extinction ratio and insertion loss. Figures 5 (d), (e), and (f) are more intuitive plots showing the 3-dB bandwidth. The EAM can easily achieve at least 10 GHz, which has potential for use in fast optical communication systems. However, under a large electric field, carriers maintain the saturation velocity, which physically limits the maximum bandwidth of the EAM. For our device with optimized dimensions, the maximum achievable bandwidth is limited to 32.6 GHz (see Supplement 1).

**Benchmarking and comparison**

Table 1. Specs comparisons for some recent EAM reports of different active layers[a]

| Active Layer | Structure | ER (dB) | IL (dB) | Switching Energy (fJ/bit) | Linear Footprint (um) | Bandwidth (GHz) | Ref. Number |
|---|---|---|---|---|---|---|---|
| ITO | Au/SiO$_2$/ITO MOS cap | 5 | 1 | 60 | 5 | \ | 5 |
| ITO | Cu/ITO/HfO$_2$/TiN | 19.9 | 2.9 | 400 | 1 | 11 | 6 |

|  | MOS cap |  |  |  |  |  |  |
|---|---|---|---|---|---|---|---|
| ITO | Coupling-enhanced dual-gate | 2 | 2 | 770 | 4 | 0.45-5.4 | 7 |
| Graphene | Graphene/hBN on top | 28 | 1.28 | 212 | 120 | 46.4 | 8 |
| Graphene | Graphene with bulk silver on top | 3.6 | 2.7 | 0.4 | 3 | >100 | 9 |
| MQWs | Ge/SiGe MQWs, p-i-n diode | 10 | 5 | 108 | 90 | 23 | 10 |
| QDs | In/As QDs, p-i-n diode | 5 | 8 | \ | 10 | \ | 11 |
| TMDs | TMD monolayers on silicon | 9 | \ | \ | 1 | \ | 12 |
| **This Work** | **$WS_2/Al_2O_3$ on slot-shaped silver** | **9.1** | **7.6** | **2.96** | **0.4** | **15** | **\** |

[a] Extinction ratio (ER) describes the ratio between the on-state transmission and off-state transmission. Insertion loss (IL) describes the loss of the light for on-state transmission. Switching energy describes the energy consumption for each 0-to-1 transition in the on-off keying modulation. 3-dB bandwidth describes the maximum switching frequency allowed. The corresponding reference numbers are listed in the last column for convenience.

In order to compare the advantages and disadvantages with recent reports of other electro-absorption modulators (EAMs) with different active layers [5–12], we have analyzed the extinction ratio, insertion loss, linear footprint, switching energy, and the 3-dB bandwidth in figure 7. EAMs with ITO active layers require MOS capacitor structures to control the injection of carriers. However, ITO-based EAMs have a large switching energy in order to achieve significant tunability due to the low light confinement of Silicon-based waveguides and a large geometric capacitance. Graphene-based EAMs are often used in super-fast interconnects, but come with a larger area. The fast speed is due to the effects of quantum capacitance, with the geometric capacitance contributing little to the energy. Multiple quantum wells (MQWs) and quantum dots (QDs) structures are difficult to predict the performance of due to different mechanisms. EAMs that adopt quantum stark-effect mechanisms can operate at higher frequencies, even in the visible frequency domain. Our proposed EAM shows a good extinction ratio, low energy consumption, an extremely small linear footprint, and a comparable bandwidth. However, due to the strong coupling phenomenon, the insertion loss of our EAM is inevitably higher than many other structures because of the lossy nature of plasmonic structures.

**Conclusions and outlook**
In conclusion we have presented a comprehensive theoretical study of an ultra-compact, plexcitonic EAM with comparable performance metrics to other plasmonic, quantum well and 2D materials based EAMs. Our proposed EAM has the smallest lateral footprint along with < 3 fJ/bit switching energy, making it a promising candidate for compact electro-optical modulator. Our theory and model suggest that in order to further improve the performance of this plexcitonic EAM, higher electric fields are necessary which may be obtained via a higher external voltage applied to the superlattice and also reducing the area of the contact region of metal to $WS_2$. Finally improving the electronic quality of the monolayer $WS_2$ and reducing metal/$WS_2$ contact resistance will also help improve the performance of this plexcitonic EAM design. Given that the EO effect in $WS_2$ is well established and superlattices used in this study have already been demonstrated, future work will focus on experimental demonstration of this concept. We believe our approach of using strong light-matter coupling and hybrid states for EAMs is promising for submicron-scale modulator design concurrently with high extinction ratio, speed, and low-power operation.

**Acknowledgements.** D.J., R.Y. and J.L. acknowledge primary support for this work by the Asian Office of Aerospace Research and Development (AOARD) and the Air Force Office of Scientific Research (AFOSR) FA2386-20-1-4074 and FA2386-21-1-4063. D.J. also acknowledges partial support from the Office of Naval Research (N00014-23-1-203) University Research Foundation at Penn and the Alfred P. Sloan Foundation for the Sloan Fellowship.

**Conflict of Interest.** The authors declare no conflicts of interest.

**Data availability.** Optical dielectric function data underlying the results presented in this paper are available in Ref. [21]. The data presented in this paper is available from the corresponding author upon reasonable request.

**Supplementary information.** A supplementary information is available which contains additional details and analysis on electrical power consumption, bandwidth calculations and geometric design optimization

# Ultra-compact plexcitonic electro-absorption modulator

Ruoyu Yuan, Jason Lynch and Deep Jariwala[*]
*Department of Electrical and Systems Engineering, University of Pennsylvania, PA, 19104, Philadelphia, USA*
\* Corresponding author: dmj@seas.upenn.edu

# Supplementary information

**Bandwidth evaluations**

To calculate the bandwidth of our electro-absorption modulator (EAM), we use the following parameters to derive the formulas.

Table S1. Parameters for bandwidth calculation of our optimized EAM.

| Parameter Symbol | Meaning | Value |
|---|---|---|
| $W_s$ | Contour Width of the device from electrode contact to the other side. | 920 $nm$ |
| $L$ | Length of the device. | 400 $nm$ |
| $S$ | The total surface area of the device. ($W_s L$) | \ |
| $U_1$ | The steady voltage across the capacitor. | 0.67 $V$ |
| $E_{eff}$ | The effective electric field driving the carriers to flow from the contact to the other side. | \ |
| $U$ | The transient voltage across the capacitor. | \ |
| $t_{setup}$ | The minimum time that is needed for uniform distribution after injecting free carriers into the WS$_2$ monolayer. | \ |
| $n_{2D}$ | 2D carrier concentration, which has the unit of $m^{-2}$. | \ |
| $v_{sat}$ | Saturation velocity of electrons in WS$_2$ monolayer. | $3 \times 10^4\ m/s$ |
| $E_{sat}$ | Saturation electric field of electrons in WS$_2$ monolayer. | $6 \times 10^5\ V/m$ |
| $\mu$ | Mobility, which is defined by $v_{sat}/E_{sat}$ (SI unit). | \ |

The carriers are injected from the electrode contact and flow to the other side of the device. In semiconductors, especially in TMD monolayers, electrons have low mobility and thus have low saturation velocity. Recent study [1] experimentally gives the saturation velocity of monolayer WS$_2$, which can be used to estimate the minimum switching time for carriers to be fully injected.

$$t_{setup} = \frac{W_s}{v_{sat}} = 3.07 \times 10^{-11} s \tag{S1}$$

This shows that the upper limit of modulation speed is 32.6 GHz in our work. To balance the accuracy and complexity when we evaluate the bandwidth, we simplify the electrical properties of electrons by a linear model in figure S1.

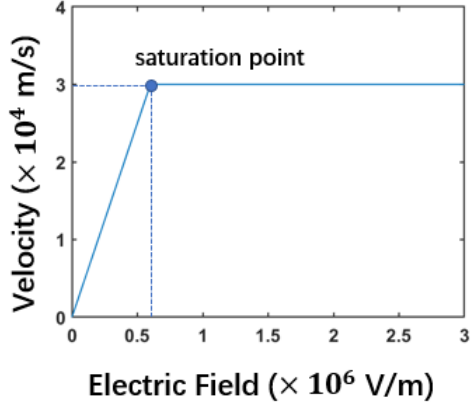

Fig. S1. Linear model describes the relationship between the velocity of an electron and the electric field applied to it. Saturation comes from damping mechanisms when carriers are injected.

The relationship between the velocity and the electric field is:
$$v = \begin{cases} \mu E, & E < E_{sat} \\ v_{sat}, & E > E_{sat} \end{cases} \quad (S2)$$

When a carrier moves from the electrode to the other side, it is driven by the voltage source. At the end of the setup time, the charge is moved $W_s$ away from the electrode and the WS$_2$ plate has the potential $U$.
$$W = q \int_l \mathbf{E} d\mathbf{l} = qE_{eff}W_s = q(U_1 - U) \quad (S3)$$

In equation (S3), we assume an effective electric field that is constantly applied to the carrier to avoid difficult integral. For monolayer WS$_2$, the current is given by
$$I(t) = n_{2D}(t)qv(t)L \quad (S4)$$

From equation (S2), (S3), and (S4), we get
$$I(t) = n_{2D}(t)qL \begin{cases} \mu \frac{U_1 - U(t)}{W_s}, & U(t) > U_1 - W_s E_{sat} \\ v_{sat}, & U(t) \leq U_1 - W_s E_{sat} \end{cases} \quad (S5)$$

The surface carrier concentration can be calculated by
$$n_{2D}(t) = \frac{CU(t)}{qS} \quad (S6)$$

where $C$ is the total capacitance induced by two WS$_2$ monolayers separated by Al$_2$O$_3$ dielectric. The capacitance effect also leads to the transient current by
$$C \frac{dU(t)}{dt} = I(t) \quad (S7)$$

Equations (S5), (S6), and (S7) form a first-order differential equation
$$\begin{cases} \frac{dU(t)}{dt} = \frac{L\mu}{SW_s} U(t)[U_1 - U(t)], & U(t) > U_1 - W_s E_{sat} \\ \frac{dU(t)}{dt} = \frac{Lv_{sat}}{S} U(t), & U(t) \leq U_1 - W_s E_{sat} \end{cases} \quad (S8)$$

General solutions of equation (S8) are
$$\begin{cases} U(t) = \frac{C_1 U_1 e^{\frac{L\mu U_1}{SW_s}t}}{C_1 e^{\frac{L\mu U_1}{SW_s}t} - 1}, & U(t) > U_1 - W_s E_{sat} \\ U(t) = e^{\frac{Lv_{sat}}{S}t} + C_2, & U(t) \leq U_1 - W_s E_{sat} \end{cases} \quad (S9)$$

where $C_1$ and $C_2$ are arbitrary numbers. With the initial condition and continuity of $U(t)$ through the entire charging process
$$\begin{cases} U(t_0) = U_1 - W_s E_{sat}, \lim_{t \to t_0^-} U(t) = \lim_{t \to t_0^+} U(t) \\ U(0) = 0 \end{cases} \quad (S10)$$

The solutions to describe the transient performance are

$$\begin{cases} U(t) = \dfrac{C_1 U_1 e^{\frac{L\mu U_1}{SW_s}t}}{C_1 e^{\frac{L\mu U_1}{SW_s}t}-1}, C_1 = \dfrac{W_s E_{sat}-U_1}{W_s E_{sat}}(U_1 - W_s E_{sat}+1)^{\frac{-\mu U_1}{W_s v_{sat}}}, t > \dfrac{S}{Lv_{sat}}\ln(U_1 - W_s E_{sat}+1) \\ U(t) = e^{\frac{Lv_{sat}}{S}t}-1,\ t \le \dfrac{S}{Lv_{sat}}\ln(U_1 - W_s E_{sat}+1) \end{cases} \quad (S11)$$

Based on equation (S11), we get the simulation results in figure 5 of the primary manuscript.

**Geometric optimization**

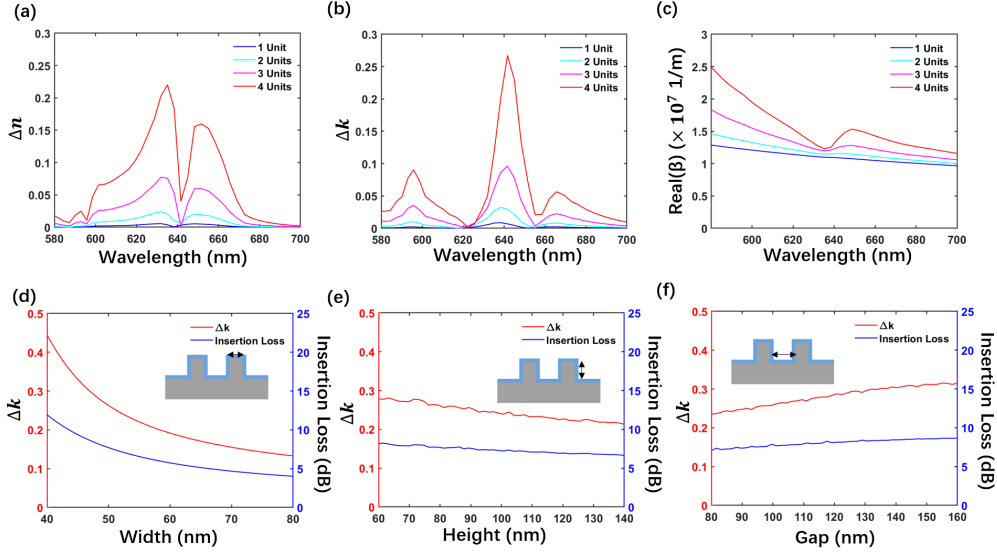

Fig. S2. (a) The n value (real part of the refractive index) vs wavelength under different units of the superlattice structure. (b) The k value (imaginary part of the refractive index) vs wavelength under different units of the superlattice structure. (c) The real part of the propagation constant vs wavelength under different units of the superlattice structure. (d) The k value and the corresponding insertion loss vs width. (e) The k value and the corresponding insertion loss vs height. (f) The k value and the corresponding insertion loss vs gap.

The optimization process demonstrates the impact of geometric parameters on the coupling parameter $g$ and the corresponding optical response. Theoretically, we can manipulate the number of excitons by increasing the number of units in a superlattice and adjust the mode volume by changing the sizes according to equation (2) in the primary manuscript. Figures S2(a), (b), and (c) illustrate the tunability at different number of excitons. The more excitons present in the system, the larger the coupling parameter will be, leading to a higher off-state absorption. However, it should be noted that the fabrication process becomes more complex as the number of units in the superlattice increases, so a design with 4 units is appropriate. Figures S2(d), (e), and (f) show the effect of changing the sizes of the EAM and adjusting the mode volume. The width of the device has a significant impact on the tunability, but it also increases the insertion loss, which is a characteristic of the SPP mode. The height and gap designs are limited by electrical properties, such as the requirement for a small geometric capacitance for energy efficiency and a short contour width for high bandwidth. In conclusion, the number of units is a trade-off between fabrication complexity, insertion loss, and tunability, while both electrical and optical characteristics must be considered in the geometric design. Our optimized EAM demonstrates a balance of these requirements. It is important to note that the tunable refractive index data may vary for different $WS_2$ monolayer samples, so optimization should always follow the principles outlined in the primary manuscript.